\documentclass[final]{aipproc}

\layoutstyle{6x9}

\def\beq{\begin{equation}}
\def\eeq{\end{equation}}
\def\vp{\varphi}
\begin{document}

\def \lta {\mathrel{\vcenter
     {\hbox{$<$}\nointerlineskip\hbox{$\sim$}}}}
\def \gta {\mathrel{\vcenter
     {\hbox{$>$}\nointerlineskip\hbox{$\sim$}}}}

\title{Dark Energy and \\ The Dark Matter Relic Abundance}

\author{Francesca Rosati}{
  address={Universit\`a di Padova, Dipartimento di Fisica ``Galileo Galilei'', \\ 
and INFN - Sezione di Padova, via Marzolo 8, 35131 Padova, Italy}
}

%

\begin{abstract}
Two mechanisms by which the Quintessence  scalar could enhance the relic 
abundance of dark matter particles are discussed. 
This effect can have an impact on supersymmetric candidates for dark matter.
\end{abstract}

\maketitle

\section{Introduction}

According to the standard paradigm, a particle species goes 
through two main regimes during the cosmological evolution. 
At early times it stays in thermal equilibrium, until the particle interaction 
rate $\Gamma$ remains larger than the expansion rate H. 
Later on, the particles will be so diluted by the expansion of the universe
that they will not interact anymore and H will overcome $\Gamma$.
The epoch at which $\Gamma=$ H is called `freeze-out', and after that time 
the number of particles per comoving volume for any given species will remain
constant. This is how cold dark matter particle relics (neutralinos, for example) 
are generated. 

As it can be easily understood, this scenario strongly depends on the evolution 
equation for H in the early universe, which is usually assumed to be 
radiation--dominated. However, as it was already noticed some 
time ago \cite{barrow-turner}, there is little or no evidence that 
before Big Bang Nuclesynthesis (BBN) it was 
necessarily so. Non-standard scenarios are then worth exploring.
In particular, if we imagine that for some time in the past
the Hubble parameter was larger than usually thought (for example, due to the 
presence of some other component, in addition to radiation), then the decoupling of 
particle species  would be anticipated, resulting in a net 
enhancement of their relic abundance.

A natural candidate for doing that is the Quintessence 
scalar, which is thought to consitute the Dark Energy fluid dominating 
the present universe. In most Quintessence models, the 
cosmological scalar is assumed to become the dominant component of the 
universe after a long period of sub-domination \cite{quint}, 
playing little or no role in the earliest epochs.
However, this has not always to be the case, as we will show in the following.
We will focus on the possibility of modifying the past evolution of 
the Hubble paramenter H, with the double aim of respecting all the post-BBN 
bounds for the expansion rate and of producing a mesurable 
enhancement of the dark matter particles relic abundance.
In particular, we will report about two possible mechanism by which the past 
dynamics of the Quintessence scalar could significantly modify the standard 
evolution of the pre-BBN universe: an early ``kination'' phase and a 
``scalar-tensor'' model.

\section{Kination enhancement}

If we imagine to add a significant fraction of scalar 
energy density to the background radiation at some time in the cosmological history, 
this would produce a variation in $H^2$, depending 
on the scalar equation of state $w_\phi$.
If $w_\phi > w_r=1/3$, the scalar energy density  
would decay more rapidly than radiation, but temporarily increase the global
expansion rate. This possibility was explicitly considered in 
Ref.~\cite{salati}, where it was calculated that a huge enhancement of 
the relic abundance of neutralinos could be produced in this way.

In a flat universe, a scalar field with potential $V(\phi)$ obeys the 
equations
\begin{equation}
\ddot{\phi} + 3H\dot{\phi} + dV/d\phi = 0  \,\,\,  ;  \,\,\,\,\,\,\,\
H^2 \equiv ( \dot{a}/a )^2 = 8\pi\rho /3M_p^2  \,\,\, .
\label{scaleq}
\end{equation}
For any given time during the cosmological evolution,
the relative importance of the scalar energy density
w.r.t.~to matter and radiation in the total energy density 
$\rho \equiv \rho_{m} + \rho_{r} + \rho_{\phi}$
depends on the initial conditions, and is constrained by the available
cosmological data on the expansion rate and large scale structure.
If the potential $V(\phi)$ is of the runaway type, the initial stage of the scalar evolution is typically 
characterized by a period of so--called `kination' \cite{quint}
during which the scalar energy density
$\rho_{\phi} \equiv \dot{\phi}^2/2 + V(\phi)$
is dominated by the kinetic contribution $E_k= \dot{\phi}^2/2 \gg V(\phi)$,
giving $w_\phi =1$.
After this initial phase, the field comes to a stop and remains nearly 
constant for some time (`freezing' phase), until it eventually reaches an 
attractor solution \cite{quint}.

Then, if we modify the standard picture 
according to which only radiation plays a role in the post-inflationary era and 
suppose that at some time $\hat{t}$  the scalar contribution was 
small but non negligible w.r.t.~radiation, then at that time the 
expansion rate $H(\hat{t})$ should be correspondingly modified.
During the kination phase the scalar
to radiation energy density ratio evolves like 
$\rho_{\phi}/\rho_r \sim a^{-3(w_\phi - w_r)} = a^{-2}$, and so
the scalar contribution would rapidly fall off and leave room to 
radiation.
In this way, we can respect the BBN bounds and at the same time keep a 
significant scalar contribution to the total energy density just few red-shifts 
before.
The increase in the expansion rate $H$ due to the additional scalar 
contribution would anticipate the decoupling of particle species and 
result in a net increase of the corresponding relic densities.
As shown in \cite{salati}, a scalar to radiation energy density ratio
$\rho_{\phi}/\rho_r \simeq 0.01$ at BBN would give an enhancement
of the neutralino codensity of roughly three orders of magnitude.

The enhancement of the relic density of neutralinos requires that at some 
early time the scalar energy density was dominating the Universe. 
This fact raises a problem if we want to identify the scalar
contribution responsible for this phenomenon with the
Quintessence field \cite{rosati03}.  Indeed, the initial conditions
must be such that the scalar energy density is sub-dominant 
at the beginning, if we want the Quintessence field to reach the cosmological
attractor in time to be responsible for the presently
observed acceleration of the expansion \cite{quint}.
For initial conditions $\rho_{\phi}\gta \rho_r$ we obtain instead an
`overshooting' behavior: the scalar field rapidly rolls down the potential
and after the kination stage remains frozen at an energy density much smaller than the critical one.
However, as shown in \cite{mpr}, more complicated dynamics are
possible if we relax the hypothesis of considering a single uncoupled scalar.  
The presence of several scalars and/or of a
small coupling with the dark matter fields could modify the
dynamics in such a way that the attractor is reached in time even if
we started in the overshooting region.

Consider a potential of the form
$V(\phi_1,\phi_2) = M^{n+4} \left(\phi_1 \phi_2\right)^{-n/2}$,
with M a constant of dimension mass.
In this case, the two fields' dynamics 
enlarges the range of possible initial conditions
for obtaining a quintessential behavior today.
This is due to the fact that the presence of more scalars allows to play with 
the initial conditions in the fields' values, while maintaining the total 
initial scalar energy density fixed. 
Doing so, it is possible to obtain a situation in which for a fixed 
$\rho_{\phi}^{in}$ in the overshooting region, if we keep initially
$\phi_1=\phi_2$ we actually produce an overshooting behavior, 
while if we choose to start with $\phi_1\not =\phi_2$ (and {\it the same} 
$\rho_{\phi}^{in}$) it is possible to reach
the attractor in time.

Suppose, instead, that the Quintessence scalar is not completely decoupled from the 
rest of the Universe.
Among the possible interactions, two interesting 
cases are the following:
\begin{equation}
V_b = b\ H^2 \phi^2 \;\;\;\;\;
\mbox{\rm or} \;\;\;\;\;
V_c = c \rho_m \phi
\label{inter}
\end{equation}
If we add $V_b$ or $V_c$ to $V=M^{n+4}\phi^{-n}$, the potential will acquire a 
(time-dependent) minimum and the scalar field will be prevented from running freely 
to infinity.
In this way, the long freezing phase that characterizes the evolution of a 
scalar field with initial conditions in the overshooting region 
can be avoided.
A more detailed discussion, together with numerical examples, can be found in 
Ref.~\cite{rosati03}.

\section{Scalar-tensor enhancement}

A different possibility arises if we consider Quintessence models in the framework 
of scalar-tensor (ST) theories of gravity (see \cite{scalar-tensor} and references 
therein).
These theories represent a natural framework in which massless scalars
may appear in the gravitational sector of the theory without being
phenomenologically dangerous, since they assume a metric coupling of
matter with the scalar field, thus ensuring the equivalence
principle and the constancy of all non-gravitational coupling
constants \cite{damour}. Moreover  a large
class of these models exhibit an attractor mechanism towards GR \cite{dam3}, that
is, the expansion of the Universe during the matter dominated era
tends to drive the scalar fields toward a state where the theory
becomes indistinguishable from GR.

ST theories of gravity are defined, in the so--called
`Jordan' frame, by the action 
\begin{equation}
S_{g}=\frac{1}{16\pi }\int d^{4}x\sqrt{-\tilde{g}}\left[ \Phi^2 \tilde{R} \, + 
4 \,\omega (\Phi ) \tilde{g}^{\mu \nu }\partial _{\mu }\Phi
\partial _{\nu }\Phi -4\tilde{V}(\Phi )\right] \,.  \label{jordan}
\end{equation}
The matter fields $\Psi _{m}$ are coupled only to the metric tensor $%
\tilde{g}_{\mu \nu }$ and not to $\Phi $, {\it i.e.} $S_{m}=S_{m}[\Psi _{m},%
\tilde{g}_{\mu \nu }]$.  Each ST model is identified by the
two functions $\omega (\Phi )$ and $\tilde{V}(\Phi )$.
The matter energy-momentum tensor is conserved, masses and
non-gravitational couplings are time independent, and in a locally
inertial frame non gravitational physics laws take their usual
form. Thus, the `Jordan' frame variables $\tilde{g}_{\mu \nu }$ and
$\Phi $ are also denoted as the `physical' ones in the literature. 
By means of a conformal transformation, 
\begin{equation}
\tilde{g}_{\mu \nu } \equiv \displaystyle A^{2}(\varphi )g_{\mu \nu } 
\;\;\; , \;\;\;\;\;\;
\Phi^2 \equiv \displaystyle 8 \pi M_{\ast}^2 A^{-2}(\varphi ) 
\label{transf}
\end{equation}
with
\begin{equation}
\alpha ^{2}(\varphi ) \equiv \displaystyle \, d\log A(\varphi )/d\varphi \,
=\, 1/(4\omega (\Phi )+6)\,,
\end{equation}
it is possible to go the `Einstein' frame
in which the gravitational action takes the standard form, while
matter couples to $\vp$ only through a purely metric coupling, 
\begin{equation} 
S_m = S_{m}[\Psi_{m},A^{2}(\varphi ){g}_{\mu \nu }] \,\,\, . 
\end{equation}
In this frame masses and non-gravitational coupling constants are
field-dependent, and the energy-momentum tensor of matter fields is
not conserved separately, but only when summed with the scalar field
one. On the other hand, the Einstein frame Planck mass $M_{\ast }$ is
time-independent and the field equations have the simple form
\begin{equation}
R_{\mu \nu }-\mbox{\small $\frac{1}{2}$}g_{\mu \nu }R =
T^\vp_{\mu \nu }/M_{\ast}^2 + T_{\mu \nu }/M_{\ast}^2  
\;\;\; , \;\;\;\;\;\;
M_{\ast}^2 \partial ^2 \varphi + \partial V/\partial \varphi  = 
- \alpha (\varphi ) T/\sqrt{2} \, .
\label{eomgen}
\end{equation}
When $\alpha(\varphi)=0$ the scalar field is
decoupled from ordinary matter and the ST theory is indistinguishable
from ordinary GR.
The effect of the early presence of a scalar field on the physical processes will 
come through 
the Jordan-frame Hubble parameter $\tilde H \equiv d \log \tilde{a}/d\tilde{\tau}$:
\begin{equation}
\tilde{H} = H \, (1 + \alpha (\varphi) \, \varphi ^{\prime})/A(\vp) \, , 
\label{Htilde}
\end{equation}
where $H\equiv d\log a/d \tau$ is the Einstein frame Hubble parameter.
A very attractive class of models is that in which the function
$\alpha(\varphi)$ has a zero with a positive slope, since this point,
corresponding to GR, is an attractive fixed point for the field
equation of motion \cite{dam3}.
It was emphasized in Ref.~\cite{max}  that the fixed point starts to be
effective around matter-radiation equivalence, and that it governs the
field evolution until recent epochs, when the Quintessence potential
becomes dominant. If the latter has a run-away behavior, the same
should be true for $\alpha(\varphi)$, so that the late-time behavior
converges to GR.
For this reason, we will consider the following choice ,
\beq A(\varphi)=1 +B e^{-\beta \varphi}\,
\;\;\; , \;\;\;\;\;\;
\alpha(\varphi) = - \beta B e^{-\beta \varphi}/(1 +B e^{-\beta \varphi})\,,
\label{alphaphi}
\eeq 
which has a run-away behavior with positive slope.

In Ref.~\cite{scalar-tensor} it was calculated the effect of ST 
on the Jordan-frame Hubble parameter $\tilde{H}$ at the time of WIMP 
decoupling, imposing on the parameters $B$ and $\beta$ the constraints 
coming from GR test, CMB observations and BBN.
Computing the ratio $\tilde{H}/\tilde{H}_{\rm GR}$ at the decoupling time of a 
typical WIMP of mass $m=200$~GeV, it was found that it is possible to
produce an enhancement of the expansion rate up to $O(10^5)$.
As a further step, it was performed the calculation of the relic abundance 
of a DM WIMP with mass $m$ and annihilation cross-section $\langle\sigma_{\rm ann}
v\rangle$.  
The effect of the modified ST gravity enters the computation of
particle physics processes (like the WIMP relic abundance) through the
``physical'' expansion rate $\tilde{H}$ defined in Eq.~(\ref{Htilde}).
We have therefore implemented the standard Boltzmann equation with
the modified physical Hubble parameter $\tilde{H}$:
\begin{equation}
dY/dx = - s \langle\sigma_{\rm ann} v\rangle (Y^2 - Y_{\rm eq}^2)/ \tilde{H}~ x
\label{eq:boltzmann}
\end{equation}
where $x=m/T$, $s=(2\pi^2/45)~h_\star(T)~T^3$ is the entropy density
and $Y=n/s$ is the WIMP density per comoving volume.
%
\begin{figure}[t] \centering
\vspace{-20pt}
\includegraphics[height=.4\textheight]{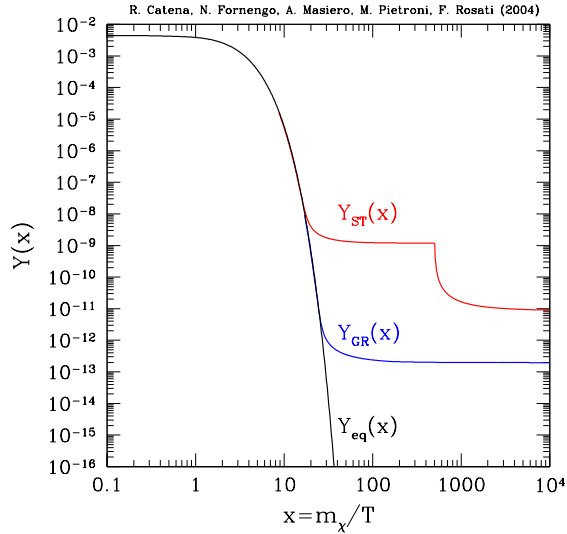}
\vspace{-20pt}
\caption{\label{fig:abundance} Numerical solution of the Boltzmann
equation Eq.~(\ref{eq:boltzmann}) in a ST cosmology for a toy--model
of a DM WIMP of mass $m=50$ GeV and constant annihilation
cross-section $\langle\sigma_{\rm ann} v\rangle = 1\times 10^{-7}$
GeV$^{-2}$. The temperature evolution of the WIMP abundance $Y(x)$
clearly shows that freeze--out is anticipated, since the expansion
rate of the Universe is largely enhanced by the presence of the scalar
field $\varphi$. At a value $x=m/T_\varphi$ a re--annihilation phase
occurs and $Y(x)$ drops to the present day value.}
\end{figure}
%

A numerical solution of the Boltzmann equation Eq.~(\ref{eq:boltzmann})
is shown in Fig.~\ref{fig:abundance} for a
toy--model of a DM WIMP of mass $m=50$ GeV and constant annihilation
cross-section $\langle\sigma_{\rm ann} v\rangle = 1\times 10^{-7}$
GeV$^{-2}$. The temperature evolution of the WIMP abundance $Y(x)$
clearly shows that freeze--out is anticipated, since the expansion
rate of the Universe is largely enhanced by the presence of the scalar
field $\varphi$. This effect is expected. However, we note that a
peculiar effect emerges: when the ST theory approached GR (a fact
which is parametrized by $A(\vp) \rightarrow 1$ at a temperature
$T_\varphi$, which in our model is 0.1 GeV), $\tilde H$ rapidly drops
below the interaction rate $\Gamma$ establishing a short period during
which the already frozen WIMPs are still abundant enough to start a
sizeable re--annihilation. This post-freeze--out ``re--annihilation
phase'' has the effect of reducing the WIMP abundance, which
nevertheless remains much larger than in the standard case 
(for further discussion on this aspect see \cite{scalar-tensor}).
%
\begin{figure}[t] \centering
\vspace{-20pt}
\includegraphics[height=.4\textheight]{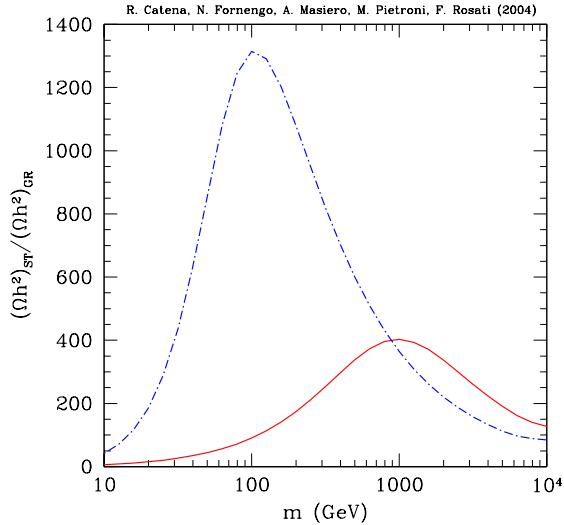}
\vspace{-20pt}
\caption{\label{fig:ratio} Increase in the WIMP relic abundance in ST
cosmology with respect to the GR case.  The solid curve refers to an
annihilation cross section constant in temperature, i.e.
$\langle\sigma_{\rm ann} v\rangle = a = 10^{-7}$ GeV$^{-2}$, while the
dashed line stands for an annihilation cross section which evolves
with temperature as $\langle\sigma_{\rm ann} v\rangle = b/x = 10^{-7}$ GeV$^{-2}/x$.}
\end{figure}
%
The amount of increase in the relic abundance which is present in ST
cosmology is shown in Fig.~\ref{fig:ratio}. The solid curve refers to
an annihilation cross section constant in temperature, i.e.
$\langle\sigma_{\rm ann} v\rangle = a$, while the dashed line stands
for an annihilation cross section which evolves with temperature as:
$\langle\sigma_{\rm ann} v\rangle = b/x$. In the case of
$s$--wave annihilation the increase in relic abundance ranges from a
factor of 10 up to a factor of 400. For a pure $b/x$ dependence, the
enhancement can be as large as 3 orders of magnitude.
Needless to say, such potentially (very) large deviations entail new 
prospects on the WIMP characterization both for the choice of the CDM 
candidates and for their direct and indirect detection probes 
\cite{scalar-tensor,profumo}.



\begin{theacknowledgments}
I would like to thank R.~Catena, N.~Fornengo, A.~Masiero and M.~Pietroni with 
whom part of the work reported here was done.
This work was partially supported by the University of Padova, research 
project n. CPDG037114.

\end{theacknowledgments}



\begin{thebibliography}{99}

\bibitem{barrow-turner}
J.~D.~Barrow, Nucl.\ Phys.\ B {\bf 208} (1982) 501;
M.~Kamionkowski and M.~S.~Turner, Phys.\ Rev.\ D {\bf 42}, 3310 (1990).


\bibitem{quint}
P.~J.~Steinhardt, L.~M.~Wang and I.~Zlatev, 1999,
Phys.\ Rev.\ D {\bf 59}, 123504;
S.~C.~Ng, N.~J.~Nunes and F.~Rosati, 2001, Phys.\ Rev.\ D {\bf 64}, 083510.

\bibitem{salati} 
P.~Salati,
Phys.\ Lett.\ B {\bf 571} (2003) 121.

\bibitem{rosati03} F.Rosati, 2003, Phys. Lett. B {\bf 570}, 5.

\bibitem{mpr} A.~Masiero, M.~Pietroni and F.~Rosati, 2000,
Phys.\ Rev.\ D {\bf 61}, 023504.

\bibitem{damour}
T.~Damour, 1996, gr-qc/9606079.

\bibitem{scalar-tensor}
R.~Catena, N.~Fornengo, A.~Masiero, M.~Pietroni and F.~Rosati,
Phys.\ Rev.\ D {\bf 70}, 063519 (2004). 

\bibitem{dam3} T.~Damour and K.~Nordtvedt, Phys. Rev. {\bf D48}, 3436 (1993);
T.~Damour and A.M.~Polyakov, Nucl.\ Phys. {\bf B423}, 532 (1994).

\bibitem{max} 
N.~Bartolo and M.~Pietroni, Phys.\ Rev.\ D {\bf 61} (2000) 023518;
S.~Matarrese, C.~Baccigalupi and F.~Perrotta, astro-ph/0403480.

\bibitem{profumo}
S.~Profumo and P.~Ullio,
JCAP {\bf 0311}, 006 (2003).

\end{thebibliography}
\end{document}